# A Two-Dimensional Approach to Evaluate the Scientific Production of Countries (Case Study: The Basic Sciences)

Ammar Nejati[1], Seyyed Mehdi Hosseini Jenab[2]

[1]*nejati_a@sharif.edu*
Sharif University of Technology, Azadi Ave, Tehran (Iran)

[2]*jenab@aut.ac.ir*
Amirkabir University of Technology, 1549, 464 Hafez Ave, Tehran (Iran)

**Abstract**
The quantity and quality of scientific output of the topmost 50 countries in the four basic sciences (agricultural & biological sciences, chemistry, mathematics, and physics & astronomy) are studied in the period of the recent 12 years (1996 – 2007). In order to rank the countries, a novel two-dimensional method is proposed, which is inspired by the H-index and other methods based on quality and quantity measures. The countries data are represented in a "quantity–quality diagram", and partitioned by a conventional statistical algorithm (*k*-means), into three clusters, members of which are rather the same in all of the basic sciences. The results offer a new perspective on the global positions of countries with regards to their scientific output.

**Introduction**

Measuring scientific production of research units (*e.g.* researchers, research groups, academic institutions, or countries) has been one of the major issues for the scientometric community, even though the offered methods and results are as important for outsiders, from academicians to policy-makers. Thus far, many measuring methods have been proposed, yet in fact, only a few have survived and remained in use; an instance of those is the Hirsch's method. Introduction of H-index [1] inspired a large amount of scientometric research focused on its applications and modifications, *e.g.* [2, 3 and 4]; yet, the core idea of the method, *i.e.* the simultaneous application of quantity and quality indices in ranking, was neither appreciated nor utilized as it deserved.

Considerable efforts have been made to evaluate scientific outputs of countries. However, in our opinion, there have been three major flaws with most of them:

(i) *Segregated indicators*. Many of the studies have inspected the relevant indicators one by one, without considering them simultaneously in one comprehensive data analysis. This practice would produce much misunderstanding, especially in science reports of nations, *e.g. SEI* report of USA [5], Japan's *NISTEP* [6] or global reports such as the *UNESCO Science Report* [7]. A usual way to overcome this problem is to include some data for other indicators besides the main one. However, even when several indicators are included, a clear and comprehensive picture is difficult to attain. To avoid this flaw, we have used a 2-dimensional data representation which includes two essential indicators simultaneously. Such a representation can provide one with a comprehensive view of the position of a country in the world of science.

(ii) *Alleged accuracy*. In many studies, exact numbers are offered as the positions of countries in the academic world. We doubt the supposed accuracy of these position markers, since usually there is an insignificant difference between countries belonging to a neighbourhood within the range of values of an indicator. To avoid this, we have clustered the countries in our 2-dimensional representation.

Previously, May [8] and King [9], in their outstanding works, have applied their two-dimensional methods to rank the countries regarding their scientific productions. Our work is in line with theirs, although our method is essentially different and the breadth of the study is much larger.

In this study, inspired by the two-dimensionality idea of Hirsch, we have devised a new method to rank the topmost fifty countries (as the largest research units) with respect to their explicit scientific output in the form of journal papers in basic sciences (agricultural and biological sciences, chemistry, mathematics, and physics & astronomy). We have included the data for a reasonable period of time (12 years, from 1996 to 2007) in a single diagram to increase the stability and reliability of the results and to reduce the temporal fluctuations considerably.

**Method**

Publication per population (PPP), as defined in Ref. [10], is commonly used as a measure of the quantity of scientific production of countries (instead of the absolute publication number). It removes the effect of population number when comparing differently populated countries. To have a meaningful comparison among the years, in our method, the same scaling is applied to all the years; *i.e.* the PPP data of a year have been divided by their world average of the same year; hence, a modification of PPP is obtained (PPPm).

Citation per publication (CPP), as defined in Ref. [10], is usually used as a measure of quality of scientific production of countries. Using CPP (instead of absolute citation number) eliminates the effect of publication number in quality comparisons. Furthermore, since it takes time for a scientific publication to be cited, CPP will decrease as approaching the final years of study and this would conceal the actual trend of the citations. To dispense with this effect, in our method, CPP data have been divided by their world average of the same year (hence, CPPm). In this manner, data for different years can be compared to each other and the temporal trends appear in this way.

The PPPm–CPPm data of the topmost fifty countries in recent twelve years (1996 – 2007) are represented as points in a single two-dimensional "quantity–quality diagram". Each point in this diagram represents the position (PPPm and CPPm) of a country at a certain year of study such that every diagram has 600 (51 times 12) points (see Fig. 1 for an example of raw data).

We have included the data for a reasonable period of time (more than a decade) in a single diagram to increase the stability and reliability of the clustering results and to reduce the temporal fluctuations considerably.

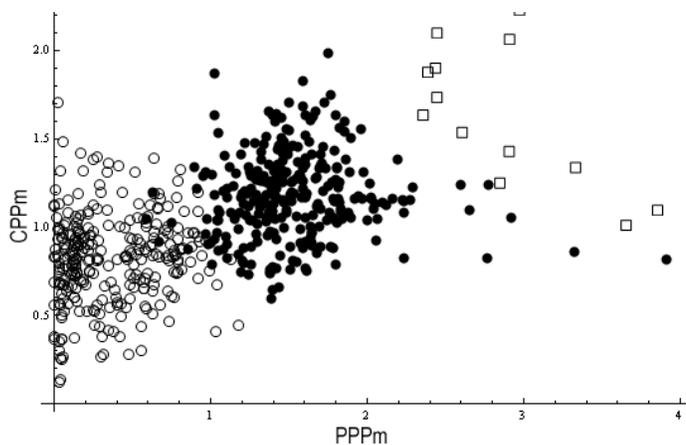

Fig. 1: A sample of raw data partitioned into 3 clusters. Each point corresponds to the PPPm–CPPm co-ordinates of a country in a certain year. These data correspond to mathematics.

At the final step, the points are clustered with the statistical clustering package of Wolfram's *Mathematica* (using *FindClusters* algorithm). The same results were obtained by the *k*-means algorithm [11, chapter 9]. An inherent property of these clustering algorithms (like *k*-means) is that the ultimate number of clusters is not fixed and should be determined externally [11]. We have decided the ultimate number of clusters to be 3 after examining other possibilities (4 or 5 clusters) after applying such a clustering to the four basic sciences, and comparing the results to find the best choice.

The data needed for the analyses have been obtained from the *SCImago* project [12] which provides Scopus data arranged according to country, branch of science and year. The population data are obtained from the *World Development Indicators* database [13].[1]

**Results and Discussion**

By comparing the clusters in the four basic sciences, certain common patterns have been observed in positioning of countries. Accordingly, three country clusters have been recognized (Figs. 2–5 and Table 1). Of course, a few countries, referred to as the "transitional countries", had different positions (cluster B or cluster C) in different basic sciences. The following are the observed clusters:

*Cluster (A):* Denmark (DK), Israel (IL), Sweden (SE) and Switzerland (CH).

*Cluster (B):* Australia (AU), Austria (AT), Belgium (BE), Canada (CA), Czech Republic (CZ), Finland (FI), France (FR), Germany (DE), Greece (GR), Hong Kong (HK), Hungary (HU), Italy (IT), Ireland (IE), Netherlands (NL), New Zealand (NZ), Norway (NO), Portugal (PT), Spain (ES), Slovenia (SI), Singapore (SG), United Kingdom (UK) and United States (US).

*Transitional countries (Tr):* Japan (JP), Slovakia (SK) and Taiwan (TW).

*Cluster (C):* Argentina (AR), Belarus (BY), Brazil (BR), Bulgaria (BG), Chile (CL), China (CN), Croatia (HR), Egypt (EG), India (IN), Iran (IR), South Korea (KR), Lithuania (LT), Malaysia (MY), Mexico (MX), Pakistan (PK), Poland (PL), Romania (RO), Russia (RU), South Africa (ZA), Turkey (TR) and Ukraine (UK).

Table 1: Countries and their corresponding clusters in each basic science.

| country codes | Agri. | Chem. | Math. | Phys. | Overall |
|---|---|---|---|---|---|
| AR | C | C | C | C | C |
| AT | B | B | B | B | B |
| AU | A | B | B | B | B |
| BE | B | B | B | B | B |
| BG | C | C | C | C | B |
| BR | C | C | C | C | C |
| BY | C | C | C | C | C |
| CA | A | B | B | B | B |
| CH | A | A | A | A | A |
| CL | C | C | C | C | C |
| CN | C | C | C | C | C |
| CZ | B | B | B | B | B |
| DE | B | B | B | B | B |
| DK | A | A | B | B | A |
| EG | C | C | C | C | C |
| ES | B | B | B | B | B |
| FI | A | B | B | B | B |
| FR | B | B | B | B | B |
| GR | B | C | B | B | B |
| HK | B | B | B | B | B |
| HR | C | C | C | C | C |
| HU | C | B | B | B | B |
| IE | B | B | B | B | B |
| IL | B | A | A | A | A |
| IN | C | C | C | C | C |
| IR | C | C | C | C | C |
| IT | B | B | B | B | B |
| JP | C | B | C | B | Tr |
| KR | C | B | C | C | C |
| LT | C | C | C | C | C |
| MX | C | C | C | C | C |
| MY | C | C | C | C | C |
| NL | B | B | B | B | B |
| NO | A | B | B | B | B |
| NZ | A | B | B | C | B |
| PK | C | C | C | C | C |
| PL | C | C | C | B | C |
| PT | B | B | B | B | B |
| RO | C | C | C | C | C |
| RU | C | C | C | C | C |
| SE | A | A | B | A | A |
| SG | B | B | B | B | B |
| SI | B | A | B | B | B |
| SK | C | B | C | B | Tr |
| TR | C | C | C | C | C |
| TW | C | B | C | B | Tr |
| UA | C | C | C | C | C |
| UK | B | B | B | B | B |
| US | B | B | B | B | B |
| ZA | C | C | C | C | C |

It seems that the countries in cluster B are those who have been able to achieve a rather acceptable level of quantity and quality of basic science production relative to their population (Fig. 2–5). Countries in cluster C are separated from those in cluster B by a gap and need to improve their production level. In between lay the "transitional countries" that are moving toward better achievements in the basic sciences and joining the countries of cluster B. The countries in cluster A are the ones which, regarding their population, have attained a rather excellent level of quality and quantity of basic science production relative to the other countries in the world, with a considerable distance from the global average.

---
[1] The population data for years 2006 and 2007 are supposed to be nearly the same as that of year 2005.

**Conclusion**

Using the tailored scientometric indicators, PPPm and CPPm, instead of the absolute numbers of publication and citation, and developing a two-dimensional method which incorporates these indicators simultaneously, led to a rather fair comparison between different countries and revealed a novel positioning of countries considering their scientific output. Nearly-similar patterns have been observed in the four basic sciences. The results differ remarkably from the common rankings.

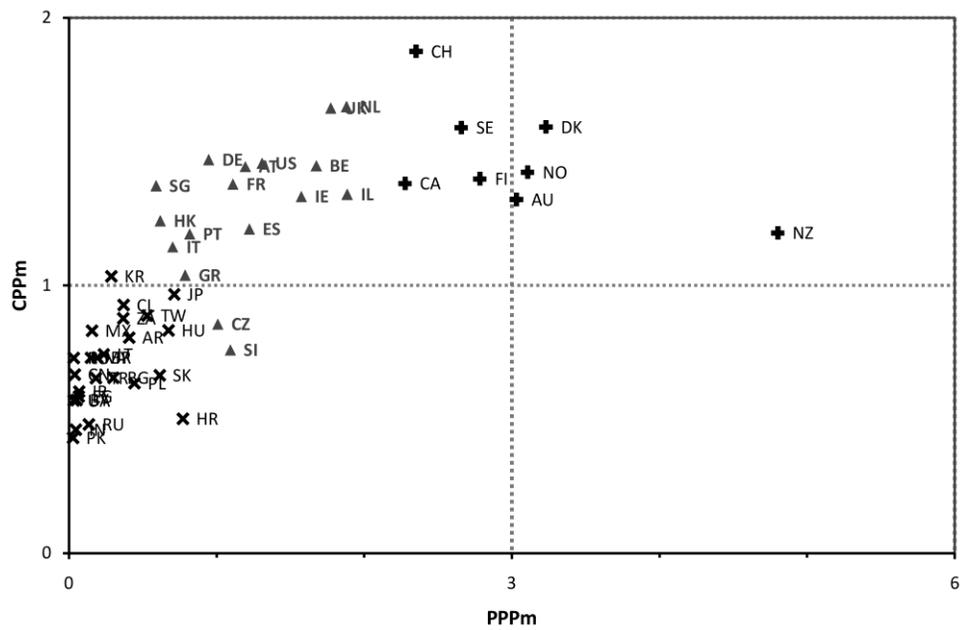

Fig. 2: Country clusters in agricultural & biological sciences.

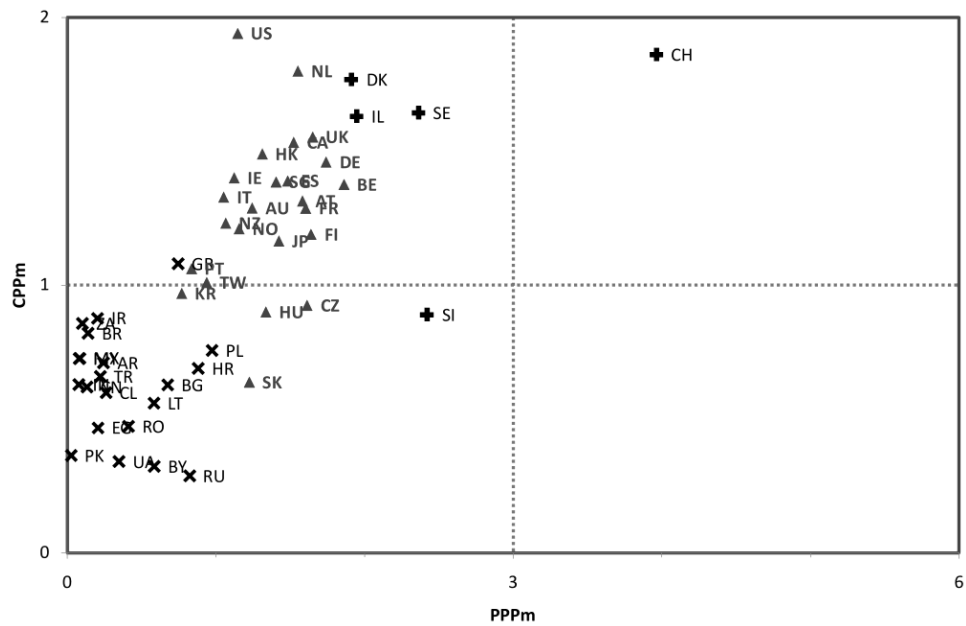

Fig. 3: Country clusters in chemistry.

Fig. 4: Country clusters in mathematics.

Fig. 2: Country clusters in physics & astronomy.